\documentclass[12pt,preprint]{emulateapj}

\usepackage{amsmath}

\begin{document}

\title{A Population of Neutron Star Ultraluminous X-ray Sources with A Helium Star Companion
}
\author{
Yong Shao$^{1,2}$, Xiang-Dong Li$^{1,2}$, and Zi-Gao Dai$^{1,2}$}

\affil{$^{1}$Department of Astronomy, Nanjing University, Nanjing 210046, People's Republic of China; shaoyong@nju.edu.cn}

\affil{$^{2}$Key laboratory of Modern Astronomy and Astrophysics (Nanjing University), Ministry of
Education, Nanjing 210046, People's Republic of China; lixd@nju.edu.cn}

\begin{abstract}

It was  recently proposed that a significant fraction of ultraluminous X-ray sources (ULXs) actually host 
a neutron star (NS) accretor. We have performed a systematic study on the NS ULX population in 
Milky Way-like galaxies, by combining binary population synthesis and detailed stellar 
evolution calculations. Besides a normal star, the ULX donor can be a helium star 
(the hydrogen envelope of its progenitor star was stripped during previous common envelope evolution) 
if the NS is accreting at a super-Eddington rate via Roche lobe overflow. 
We find that the NS$-$helium star binaries can significantly contribute the ULX population, with 
the overall number of about several in a Milky Way-like galaxy. 
Our calculations show that such ULXs are 
generally close systems with orbital period distribution peaked at $ \sim 0.1  $  day
(with a tail up to $ \sim100 $ days), and the helium stars have relatively low masses distributing 
with a maximum probability at  $ \sim 1M_{\odot} $. 

\end{abstract}


\section{Introduction}

An ultraluminous X-ray source (ULX) is an off-nuclear point-like source whose X-ray
luminosity exceeds $ 10^{39} \,\rm erg\,s^{-1}$ \citep[see][for a review]{kfr17}. 
It is generally believed that such a source is an X-ray binary (XRB) powered by accretion 
onto a compact object. An intermediate-mass ($ 10^{2}-10^{5} M_{\odot}$) 
black hole (BH) with sub-Eddington accretion was firstly suggested  by \citet{cm99} 
to account for the observed X-ray luminosities. Several ULXs with extremely high 
luminosities, peaked at $ \gtrsim 10^{41} \,\rm erg\,s^{-1}$, are thought to be promising 
intermediate-mass BH candidates \citep{f09,p14}. Alternatively the accretor is a 
stellar-mass compact object, accreting at a super-Eddington rate, as suggested by  
theoretical models \citep{k01,b02,p07} and observational features \citep{g09,s13,w18}. 
The mass of the compact object in M101 ULX-1 was dynamically measured to be 
in the stellar-mass BH range \citep{liu13}. 

\begin{table*}
\begin{center}
\caption{Basic properties of the ULX systems with an NS accretor, including peak X-ray luminosity $ L_{\rm peak} $,
pulsar's spin period $ P_{\rm spin} $, binary orbital period $ P_{\rm orb} $, donor's mass function $ f_M $, 
and estimated donor mass $ M_{\rm d} $.
\label{tbl-1}}
\begin{tabular}{lccccccl}
\\
\hline
Name   & $ L_{\rm peak} $&  $ P_{\rm spin} $ & $P_{\rm orb} $& $ f_M $& $ M_{\rm d} $\\
             &  ($\rm erg\,s^{-1}$) & (s) &  (days)&  ($M_{\odot}$) &  ($M_{\odot}$) \\
\hline
M82 X-2 (1)     &  $ 2\times 10^{40} $    & 1.37  & 2.51 (?) & 2.1 &  $\gtrsim 5.2$  \\
NGC 7793 P13 (2)   &  $  10^{40} $          & 0.42   & 63.9 & & 18$-$23  \\
NGC 5907 ULX-1 (3)     & $  10^{41} $     &  1.13     & 5.3 (?)&$ 6\times 10^{-4} $ &  \\
NGC 300 ULX-1 (4) & $ 5\times 10^{39} $ & $ \sim 31.5 $ & &$ \lesssim 8 \times 10^{-4} $ & \\
M51 ULX-8$^{*}$ (5) & $ 2\times 10^{39} $  & \\
NGC 1313 X-2 (6) & $ \sim 10^{40} $ & $ \sim 1.5 $ & $<4$ (?)& &$\lesssim 12 $ \\
M51 ULX-7 (7)   &  $7\times  10^{39} $ & 2.8 & 2 & 6.1 & $ \gtrsim 8 $ \\
\hline
\end{tabular}
\end{center}
$ ^{*} $ The compact object in the source M51 ULX-8 is believed to be an NS, since the detection of a likely
cyclotron resonance scattering feature that produced by the NS's surface magnetic field. \\
References. (1) \citet{b14}. (2) \citet{f16,i17a,m14}. (3) \citet{i17b}. (4) \citet{c18,bl18}. (5) \citet{bh18}. 
(6) \citet{sr19}. (7) \citet{rc19}.
\end{table*}

M82 X-2 is the first confirmed ULX 
hosting an accreting neutron star (NS) due to the discovery of X-ray 
pulsations \citep{b14}. To date, quite a few of ULXs have been identified to host an 
NS accretor (see Table 1). 
In some cases, Be XRBs containing an NS can appear as ULX systems, 
as their peak X-ray luminosities\footnote{Throughout this paper, 
the X-ray luminosity means the inferred luminosity for an assumed isotropic emission, even though
the emission is not actually isotropic.} reach above $ 10^{39} \,\rm erg\,s^{-1}$ 
during outbursts \citep[e.g.,][]{td17,w17,dts18}. Theoretical models indicated that
many unpulsed ULXs must actually contain an NS, because the ULX systems can be 
observed as pulsars only under rather special conditions (e.g., high spin-up rates) for the 
rotating NSs \citep{kl16,klk17}. Binary population synthesis (BPS) 
calculations showed that a large fraction of ULXs are likely to host an NS rather than
a BH accretor \citep[e.g.][]{sl15,f15,ws17}.

Until now, the properties of NS ULXs are still unclear. The detected X-ray luminosities can 
reach  $ \sim 10^{40}-10^{41}\rm \,erg$ $\rm s^{-1} $ \citep[e.g.,][]{b14,f16,i17a,i17b}, 
meaning that the NS is accreting material at a rate of $ 2-3 $ orders of magnitude higher 
than its Eddington limit. These ULXs are highly variable sources, the luminosities in the faint 
phase can drop as low as $ \lesssim 10^{37}-10^{38}\rm \,erg$ $\rm s^{-1} $. This significant 
variability is proposed to be related to the interaction between the accreting material and 
the NS's magnetic fields which were estimated to be
$ \sim 10^{9}-10^{15} $ G \citep{b14,e15,d15,kl15,t15,km16,tm16,c17,xl17}.

The nature of the donor stars in NS ULXs is not very clear.
Based on the optical observations, the ULX donor of NGC 7793 P13 was determined to be 
a BI9a star with a mass of $ 18-23M_{\odot} $ \citep{m14}. The donor mass of M82 X-2
was estimated to be greater than $ 5.2M_{\odot} $ if assuming a $ 1.4M_{\odot} $ NS 
companion \citep{b14}. For the donor of NGC 5907 ULX-1, \citet{i17b} suggested that
it is likely to be a less-evolved massive star or a less-massive (super)giant. The optical
observations with \textit{HST} data still cannot confirm the donor nature of this ULX 
system \citep{h19}. The donor mass of NGC 1313 X-2 was limited to be
$ \lesssim 12M_\odot $, provided that it is associated with a young star cluster in its vicinity \citep{sr19}.
In the source M51 ULX-7, \citet{rc19} suggested that the donor star is an
OB giant with mass $ \gtrsim 8M_\odot $. Several investigations were performed to search for the ULX donors,
but only a handful have been detected among hundreds of known ULXs \citep{r08,g13,h14}.
As a result of the observational bias that favoring bright stars, the detected donors in 
NS ULX systems tend to be luminous massive stars.

It is known that NS ULXs are binary systems in which the NS is being fed by the donor via 
Roche lobe overflow (RLOF) \citep[e.g.,][]{sl15,f15}. Modeling the evolution of NS X-ray binaries
reveals that the mass transfer is dynamically unstable when the initial mass ratio of
the donor star to the NS is larger than $ \sim 3.5 $ \citep{kolb00,pr00,t00,prp02,sl12}. 
Such a binary will go into
a common envelope \citep[CE, see a review by][]{i13} phase, and the
remnant system may be an NS$-$helium star binary if the progenitor donor has evolved off 
main-sequence prior to mass transfer \citep{bv91}. The subsequent evolution of this
binary can lead to the formation of a close system with a massive white dwarf (WD) orbited by a (partially) 
recycled pulsar  \citep[e.g.][]{vt84,dewi02,t11}. Binary evolution simulations reveal that the NS$-$helium star binaries are 
potential ULXs since the mass transfer can proceed at super-Eddington rates via RLOF \citep{ws15,tlp15}. 
However, there is little attention on the formation and evolution of these ULX binaries in the literature.
In this paper, we attempt to explore the properties of the NS ULXs with a helium star 
companion in Milky Way-like galaxies, including the parameter distribution of the binary systems 
and the number size of this ULX population. For comparison, we simultaneously provide
the information of the NS ULXs with a normal star companion \citep[see also][]{sl15}.  

The remainder of this paper is organized as follows. In Section 2 we introduce the adopted
methods, using the BPS code \textit{BSE} \citep{h02} to obtain the birthrate
distribution of incipient NS binaries and the stellar evolution code \textit{MESA} \citep{p11}
to model the binary evolutionary tracks. In Section 3, we present the calculated results and give
some discussions. Finally we make a brief summary in Section~4.

\section{Methods and Calculations}

\subsection{Generation of incipient NS binaries}

An incipient NS binary is defined as a binary system containing either a normal star 
(NS$-$normal star) just after the NS formation, or a helium star (NS$-$helium star) 
just after the CE evolution during which its hydrogen envelope is stripped by the NS. 
The subsequent evolution of incipient NS binaries may become ULX systems
if the donor supplies its material to the NS at a super-Eddington RLOF rate. To obtain the 
birthrate distributions of the incipient NS binaries, we adopt the BPS code \textit{BSE}
originally developed by \citet{h02}. With \textit{BSE} we can simulate the evolution of a large number 
of binary stars with different initial parameters, i.e. the component masses and the orbital parameters. 
The evolutionary process is assumed to begin from a primordial binary containing two zero-age 
main-sequence stars. Modeling the evolution of a binary system is then subject to many factors, 
e.g. tides, stellar winds, mass and angular momentum transfer, asymmetric supernova (SN) explosions 
and natal kicks, and CE evolution. Some modifications in the code have been made
by \citet{sl14},  in the following we will introduce some key points.

During the evolution of a primordial binary, the primary star firstly evolves to fill its Roche lobe
and supplies its envelope matter to the secondary star. If the secondary star accretes so rapidly that
it gets out of thermal equilibrium and significantly expands to fill its own Roche lobe,
then the binary goes into a contact phase \citep{ne01}. Therefore the mass transfer efficiency 
(the fraction of matter accreted onto the secondary star among the transferred matter) is an important 
factor that determining whether the primordial binary goes into a contact phase.  When dealing with the evolution
of the primordial binaries, \citet{sl14} built three mass transfer models with significantly different efficiencies. 
It is found that the rotation-dependent mass transfer model (in which the efficiency is assumed to be dependent on 
the rotational velocity of the secondary star) appears to be consistent with the observed 
parameter distributions of Galactic binaries including Be$-$BH systems \citep{sl14}, Wolf Rayet$-$O systems 
\citep{sl16} and NS$-$NS systems \citep{sl18}. So we employ the rotation-dependent mass 
transfer model in our calculations. In this model, the accretion rate onto the rotating secondary 
is assumed to be the mass transfer rate multiplying a factor of ($ 1-\Omega/\Omega_{\rm cr} $), 
where $ \Omega $ is the angular velocity of the  secondary star 
and $ \Omega_{\rm cr} $ is its critical value \citep{plv05,dm09,se09}. Since accretion of a small amount of
mass can accelerate the secondary to reach its critical rotation \citep{p81}, 
the mass transfer efficiency can be as low as $ < 0.2 $. As a consequence,
the maximal initial mass ratio of the primary to the secondary stars for avoiding the contact phase 
can reach $ \sim 6 $, and then a large number of the primordial binaries can experience stable mass transfer phases
until the primary's envelope is completely exhausted. 
The contact binaries are assumed to enter a CE phase if the primary star has evolved off main-sequence, 
otherwise they are assumed to merge into a single star \citep[see e.g.,][]{d13,sl14}. 

If a binary system goes into the CE evolution, we use the 
standard energy conservation equation \citep{w84} to calculate the orbital decay during 
the spiral-in phase.  The orbital energy of the embedded binary is used to expel the envelope.
After CE evolution, the binary system is assumed to merge into a single star
if the final separation leads to contact between the binary components, which will not contribute the ULX population.
In the code, we use the results of \citet{xl10} and \citet{w16} for the binding energy parameter 
of the donor envelope and take the CE efficiency to be 1.0\footnote{\citet{dt00} indicated that observations of the 
NS$-$WD binaries originating from a CE evolution are consistent with the CE efficiency 
of $ \lesssim 1$.  When dealing with the post-CE binaries with a WD and a main-sequence star, \citet{zs10}
suggested that the CE efficiency should be in the range of $ 0.2-0.3 $. If we adopt
a low efficiency of 0.3 instead of 1.0 in the BPS calculations, the obtained birthrates of incipient 
NS$-$helium star (NS$-$normal star) 
binaries will be decreased by a factor of $\sim 0.6  $ ($\sim 0.8  $), and the corresponding ULX 
numbers will be reduced by a factor of $\sim 0.3  $ ($\sim 0.6  $).}.
Alternatively if the mass transfer in a binary 
proceeds stably without involving a CE phase, the donor envelope will be gradually stripped via RLOF. 
In the case of stable mass transfer, we simulate the binary orbital evolution
by assuming that the ejected matter takes away the specific orbital angular momentum of the accretor. 
During the whole evolution, we use the prescription of \citet{h00} to treat the stellar wind mass losses, 
except for hot OB stars, for which we adopt the mass loss rates of \citet{v01}.  

\begin{figure*}[hbtp]
\centering
\includegraphics[width=0.6\textwidth]{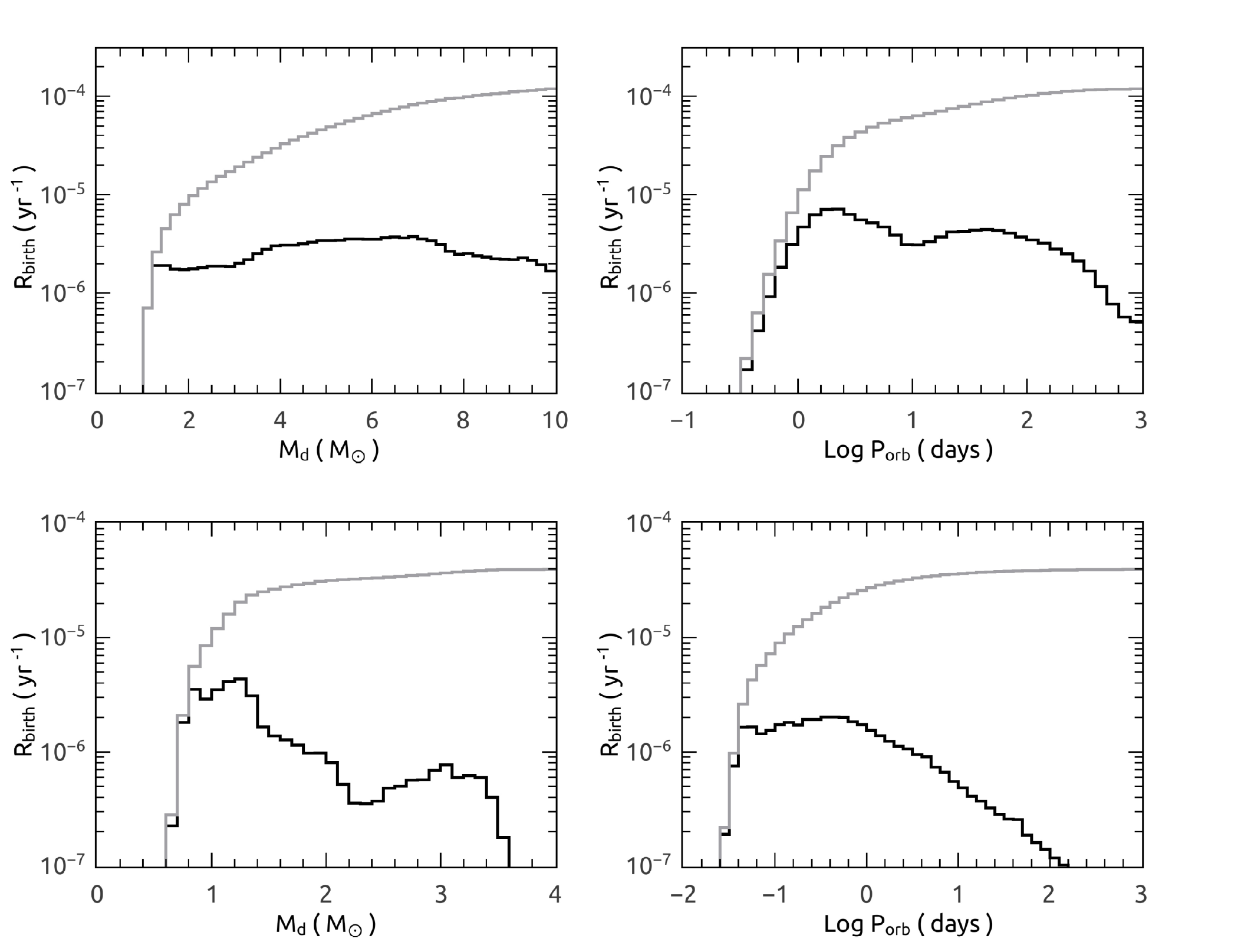}
\caption{The birthrate distributions of the incipient NS binaries with a normal star (top panels)
and a helium star (bottom panels) companion. The left and right panels depict the distributions
of the donor mass and the orbital period, respectively. In each panel, the black and grey 
curves correspond to the differential and cumulative distributions, respectively. 
   \label{figure1}}

\end{figure*}

\begin{figure*}[hbtp]
\centering
\includegraphics[width=0.6\textwidth]{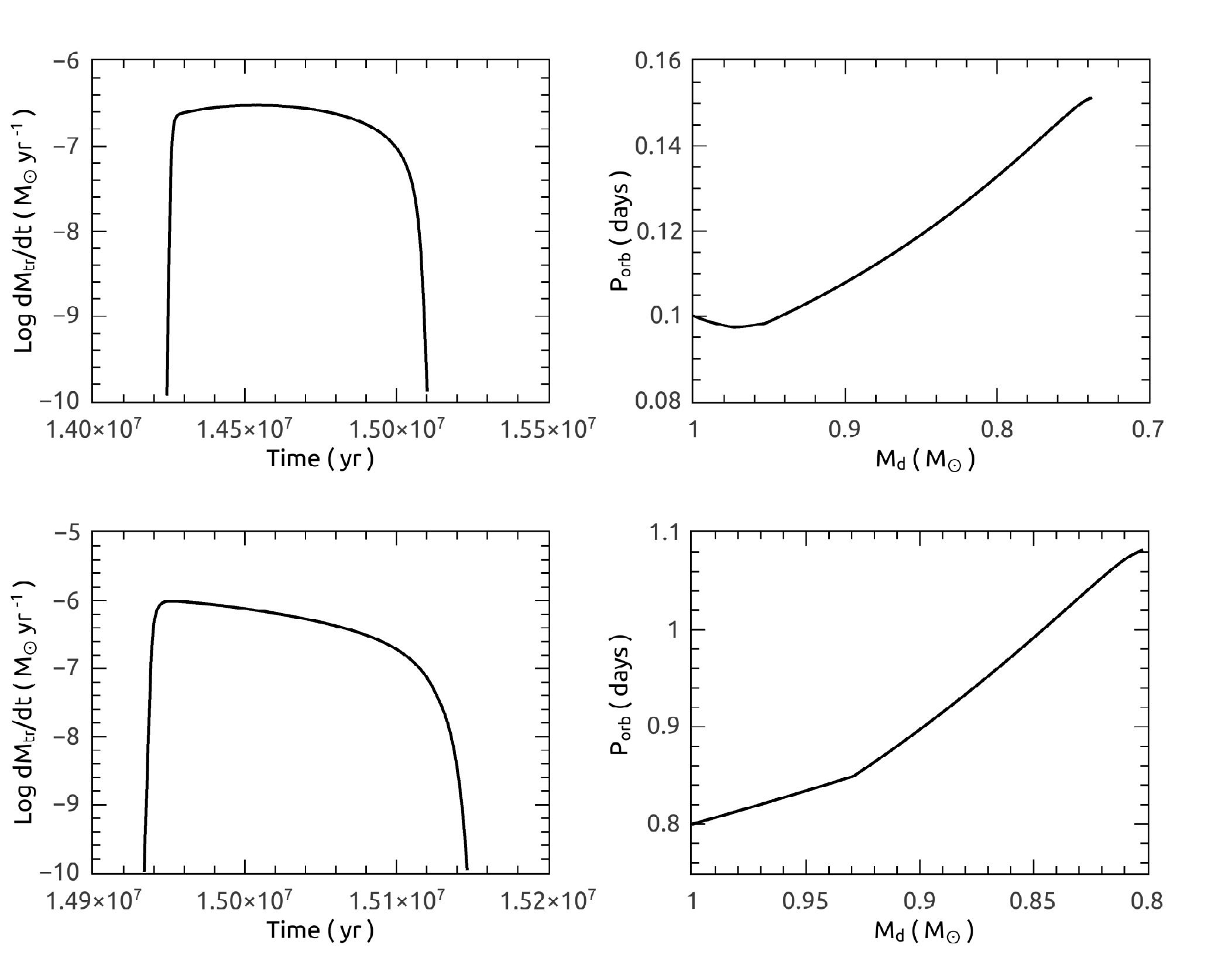}
\caption{Exampled evolution of the mass transfer rate (left panels) and the orbital period (right panels)
for two typical NS$-$helium star binaries, as a function of the time and the donor mass. The 
initial masses of binary components are $ M_{\rm NS} = 1.4M_{\odot}$ and $ M_{\rm d} = 1.0M_{\odot}$.
The top and bottom panels correspond to the initial orbital periods of 0.1 and 0.8 day, respectively.
   \label{figure1}}

\end{figure*}

We assume that NSs are formed through either electron-capture or core-collapse SNe,
the criterion suggested by \citet{f12} is used to distinguish them. In the \textit{BSE} code,
the helium core mass at the AGB base is used to set the limits for the formation of various CO cores
\citep{h02}. If the helium core mass is smaller than $ 1.83M_\odot $, the star forms a degenerate CO core,
and eventually leaves a CO WD. If the core is more massive than $ 2.25M_\odot $, the star forms
a non-degenerate CO core, stable nuclear burning will continue until the occurrence of a core-collapse SN. 
Stars with core masses between $ 1.83M_\odot $ and $ 2.25M_\odot $ form partially degenerate CO cores. 
If such a core reaches a critical mass of $ 1.08M_\odot $, it will non-explosively burn into an ONe core. 
If in subsequent evolution the ONe core can increase its mass to $ 1.38M_\odot $, the core is believed to collapse 
into an NS through an electron-capture SN.  It should be noted that the trigger of SN explosions
is actually subject to big uncertainties, and binary evolution makes it more complicated \citep[e.g.,][]{plp04}. 
During a SN explosion, the newborn NS will be imparted a natal kick, resulting in an eccentric orbit or even
disruption of the binary system. 
The kick velocities are assumed to obey the Maxwellian distributions with 
a dispersion of $\sigma = 40 \rm\, km\, s^{-1} $ \citep{d06} for NSs formed from electron-capture SNe
and $\sigma = 265 \rm\, km\, s^{-1} $ \citep{h05} for NSs formed from core-collapse SNe.

The initial parameters of the primordial binaries are taken as follows. 
The primary stars obey the initial mass function suggested by \citet{k93}, and the
mass ratios of the secondary to the primary are drawn from a flat distribution between 0 and 1. 
The orbital separations are assumed to be uniform in the logarithm \citep{a83}. We assume the initial 
orbits of all binaries are circular, as shown by \citet{h02}, the outcome of the interactions of systems with the
same semilatus rectum is almost independent of eccentricity. We adopt a binary 
fraction of 0.5 for stars with initial masses below $ 10M_{\odot} $,  otherwise we assume all massive
stars are in binaries.  The initial metallicity of stars is set to be 0.02.
We take the star formation history of Milky Way-like galaxies into account, assuming a constant 
star formation rate of $ 3 M_{\odot}\,\rm yr^{-1} $ over the past 10 Gyr period.

In Figure~1 we plot the birthrate distributions of incipient NS$-$normal star
(top panels) and NS$-$helium star (bottom panels) binaries. The left and right panels correspond to
the distributions of the donor mass and the orbital period, respectively. In each panel, the black and grey 
curves represent the differential and cumulative distributions, respectively. We can see that the total birthrate
of incipient NS$-$normal star systems is  $\sim 1.2\times 10^{-4} \,\rm yr^{-1}$.
Such incipient binaries tend to have eccentric orbits due to mass loss and kick during SN. For simplicity, we
assume that they are quickly circularized by tidal torques with the orbital angular momentum
conserved\footnote{There is a caveat that this assumption is not valid for 
long-period (e.g., $ > 20 $ days) systems. But
under this assumption, we need only to take into account the binary 
parameters of the donor masses and the orbital periods when simulating the subsequent evolution of the incipient 
NS$-$normal star binaries (see Section 2.2 below).}.
The corresponding orbital separation is then reduced by a factor of ($ 1- e^{2} $), where $ e $ is the eccentricity. 
It can be seen that the orbital period distribution
has two peaks at $ \sim 2 $ and $ \sim 50 $ days, which reflects whether the evolution of the primordial binaries 
has experienced a CE phase. The incipient NS$-$helium star binaries, as the descendants of  
long-period NS$-$normal star systems that are followed by a CE phase\footnote{It was proposed that 
an ONe WD may collapse into an NS induced by mass accretion in originally WD-helium star binaries 
\citep{chen11,liu18}. This channel may also lead to the formation of the NS$-$helium star binaries,
but is not considered in our calculations. }, have a lower birthrate of $ \sim 4 \times 10^{-5}
 \,\rm yr^{-1}$. These binaries are mainly close systems in a circular orbit, most of them have 
 orbital periods of $ \lesssim 1  $ day. 

\subsection{Evolution of NS XRBs} 

\begin{figure*}[hbtp]
\centering
\includegraphics[width=0.70\textwidth]{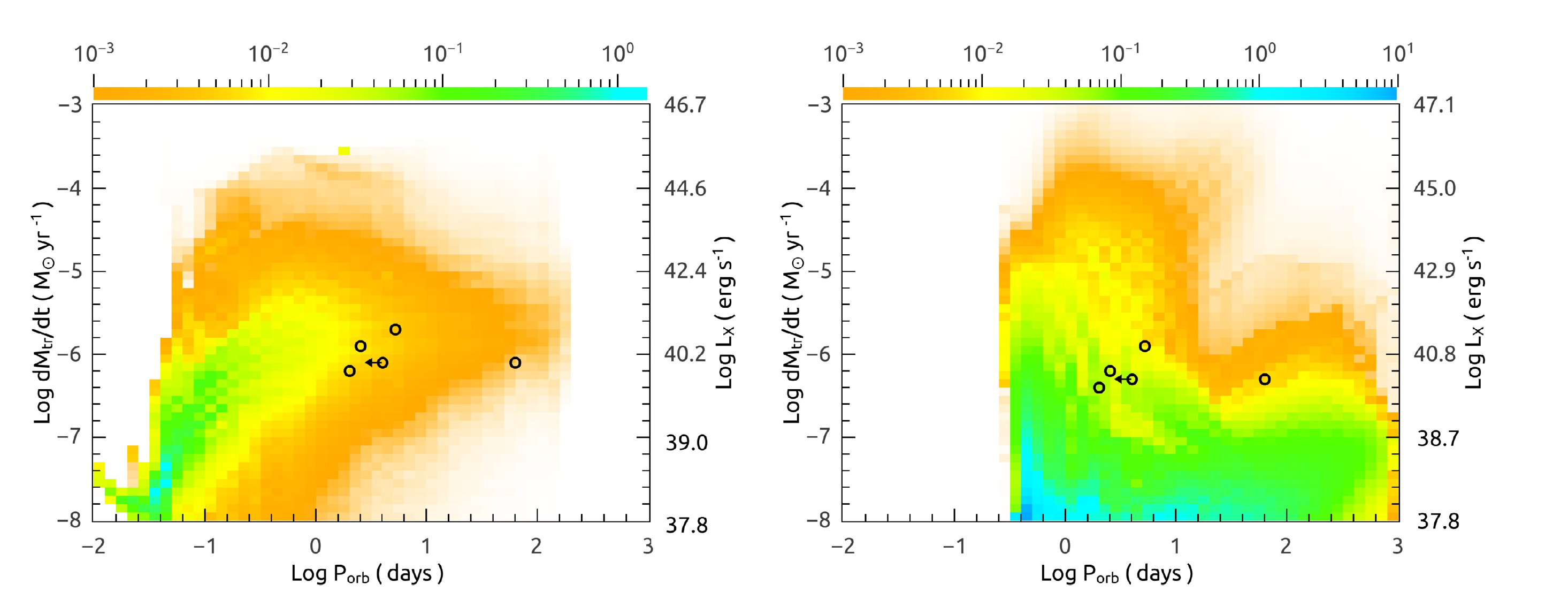}
\caption{The number distributions of NS XRBs in the orbital period $ P_{\rm orb} $ vs. mass transfer 
rate $ \dot{M} _{\rm tr}$ plane, with the assumption of a constant star formation rate 
of $ 3 M_{\odot}\,\rm yr^{-1} $ 
over a period of 10 Gyr. The left and right panels correspond to the binaries containing a helium
star and a normal star donor, respectively. Only the systems with mass transfer rates larger than $10^{-8}\,\rm M_{\odot}\,yr^{-1}$ are presented,  and the colors are scaled according to the number of the XRBs.
The five black circles denote the observed NS ULXs with known orbital periods (see Table 1).
   \label{figure1}}

\end{figure*}

Based on the obtained results in Figure~1, we track the evolutionary 
paths of incipient NS binaries with the stellar evolution code \textit{MESA} \citep[version number 10398,][]{p11,p13,p15}. 
The NS is treated as a point mass and its initial mass is set to be $ 1.4M_{\odot} $. The initial chemical compositions are 
taken to be $ X = 0.7 $, $ Y = 0.28 $, $ Z = 0.02 $ for normal stars and $ Y = 0.98 $, $ Z = 0.02 $ for helium stars. 
Each incipient binary is characterized by the donor 
mass $  M_{\rm d} $ and the orbital period $ P_{\rm orb} $. 
The incipient NS binaries from our BPS calculations are used to guide the limits of the \textit{MESA} grid 
of initial binary parameters, thus we have evolved thousands of binary systems
with different donor masses and orbital periods. For the NS$-$normal star systems, we vary the 
normal star masses from 1 to $ 10M_{\odot} $\footnote{Note that
the NS binaries with donor masses larger than $ 10M_{\odot} $ would have very small contribution to the ULX 
population in Milky Way-like galaxies \citep{sl15}.}
by steps of $ 0.1M_{\odot} $ and the orbital periods (in units of days) logarithmically from
$ -0.5 $ to 3 by steps of 0.1. For the NS$-$helium star systems, we increase the helium star masses
from 0.6 to $ 4M_{\odot} $ by steps of $ 0.1M_{\odot} $ and the orbital periods (in units of days)
logarithmically from $ -0.6 $ to 2 by steps of 0.1. These binaries are used to represent all incipient NS binaries, the 
birthrate of a specific binary is obtained by summing the ones of the incipient binaries 
reside in the corresponding grid interval of $ \Delta M_{\rm d}\times \Delta (\log P_{\rm orb}) $.

During the evolution, we adopt the scheme of \citet{r88} to compute the mass transfer rate via RLOF. We assume
the mass increase onto an NS is limited by the Eddington accretion rate ($ \sim1.5\times10^{-8} \, M_{\odot}\rm\,yr^{-1}$
for hydrogen accretion and $ \sim 4\times10^{-8} \, M_{\odot}\rm\,yr^{-1}$ for helium accretion). 
For the matter that is not accreted by the NS, we assume it escapes the binary system 
in the form of isotropic wind, taking away the NS's specific orbital angular momentum.  In some cases, the mass
transfer rates rapidly increase to $ \gtrsim 10^{-2} \,M_{\odot}\rm \,yr^{-1}$, and the code fails to converge. 
These systems are expected to go into CE evolution soon. So we use this rate as a limited condition
to judge whether the code is terminated. Whereas the ULX population are actually dominated by the binaries  
undergoing dynamically stable mass transfer with a modest rate of  $ \sim 10^{-7}  - 10^{-6}\, M_{\odot}\rm
\,yr^{-1}$, the systems with extremely high mass transfer rates cannot significantly contribute the ULX systems 
\citep[e.g.,][]{sl15}.

For the NS$-$helium star binaries, the mass transfer takes place on the nuclear timescale if the 
initial helium stars are less massive than $ \sim2.0-2.5M_\odot $, otherwise the phase of mass transfer
proceeds rapidly on the thermal timescale \citep[see also][]{dewi02}. Hence the majority of the  
NS$-$helium star binaries obtained from our BPS calculations will experience a mass transfer phase that 
is driven by the nuclear evolutionary expansion of the helium stars.
In Figure~2 we present the evolutionary tracks of two typical NS$-$helium star binaries as example.  
The initial systems contain a $ 1 M_{\odot} $ helium star around the NS, and the orbital periods are 
chosen to be 0.1 (top panels) and 0.8 day (bottom panels). In the top panels, the beginning of RLOF occurs
at the time of about $ 14.2\,\rm Myr $. The mass transfer can persist over $  0.7 \,\rm Myr $ at a
rate of  $\gtrsim 10^{-7}\, M_{\odot}\rm\,yr^{-1}$. After $ 0.26M_{\odot} $ of the envelope matter 
is transferred, the binary leaves an NS$-$WD system in a 0.15 day orbit. The final merger
of this binary will happen in $ \sim 40  \,\rm Myr$ later. In the bottom panels, the helium star evolves to fill
its Roche lobe at the time of about $ 14.9\,\rm Myr $. The mass transfer rapidly increases to 
$ \sim 10^{-6}\, M_{\odot}\rm\,yr^{-1}$ and then gradually decreases to 
$ \sim 10^{-7}\, M_{\odot}\rm\,yr^{-1}$ within a span of about $  0.2\,\rm Myr$. About $ 0.2M_{\odot} $
material is stripped during the mass transfer phase, the helium star turns into a massive WD of mass $\sim 0.8M_{\odot} $.
The binary eventually becomes an NS$-$WD binary with the orbital period of 1.08 day, which is 
similar to a system like PSR B0655+64 \citep{vt84}. It is obvious that the NS$-$helium star binaries
with typical initial parameters can spend a few tenths of $\rm Myr$ in the ULX phases.

\begin{figure*}[hbtp]
\centering
\includegraphics[width=0.7\textwidth]{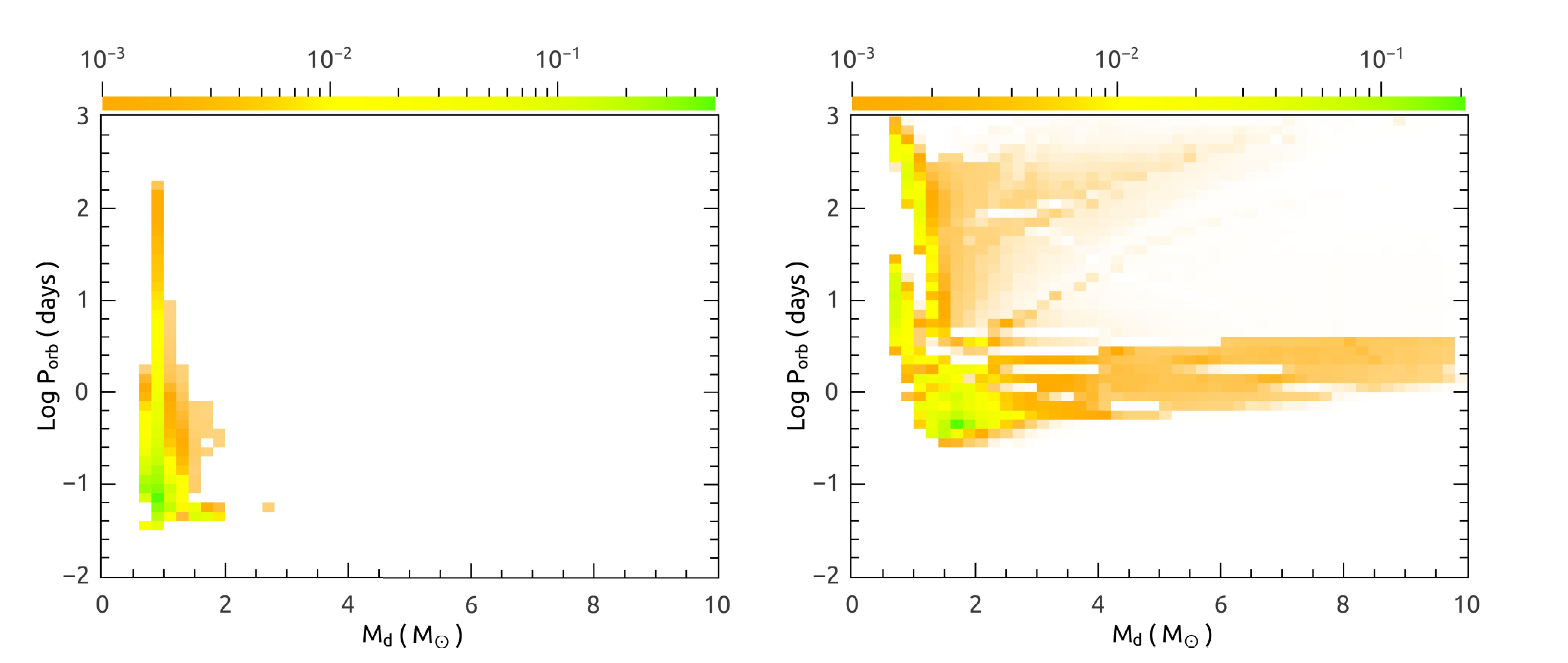}
\caption{Expected number distributions of the NS ULXs containing a helium star (left panel) or a normal star
(right panel) in the $M_{\rm d}-P_{\rm orb} $ plane.
The colors are scaled according to the number of the ULX systems. 
   \label{figure1}}

\end{figure*}

\begin{figure*}[hbtp]
\centering
\includegraphics[width=0.7\textwidth]{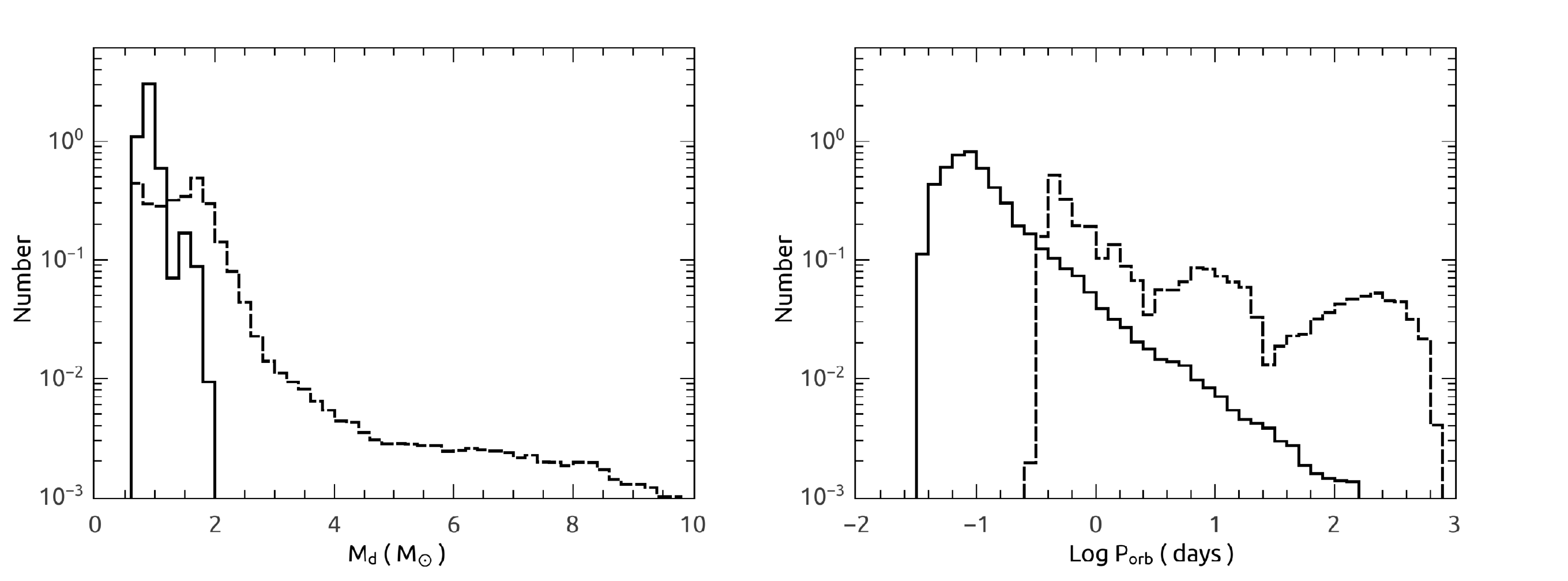}
\caption{Expected number distributions of the NS ULXs as a function of the donor mass (left panel) and the orbital
period (right panel). The solid and dashed curves denote the ULX systems containing a helium star and a normal 
star donor, respectively. 
   \label{figure1}}

\end{figure*}

Given the mass transfer rate $\dot{M} $ in an XRB, 
the X-ray luminosity can be simply estimated with
the traditional formula
\begin{equation}
L_{\rm X}=0.1\dot{M}c^{2},
\end{equation}
where $ c $ is the speed of light in vacuum. However, when $\dot{M} $ is greater than the Eddington 
rate $ \dot{M}_{\rm E} $,
the accretion disk becomes geometrically thick, which influences the X-ray luminosity. An NS that is being 
fed at a super-Eddington rate can shed more and more of the material as it approaches the NS along the disk,
thereby never violating the Eddington limit locally.
In this case, we follow \citet{kl16} to convert the mass transfer rate into the X-ray luminosity. 
The total accretion luminosity, by integrating the local disk emission \citep{ss73}, is given as 
\begin{equation}
L_{\rm acc} \simeq L_{\rm E}\left[1+\ln \left(
\frac{\dot{M}}{\dot{M}_{\rm E}} \right) \right],
\end{equation}
where $ L_{\rm E} $ is the Eddington luminosity. With this equation, the binary system can emit an 
X-ray luminosity that is limited to a few times the Eddington limit.
Due to the geometric collimation, one can see the source in
directions within one of the radiation cones, with the apparent (isotropic) X-ray 
luminosity
\begin{equation}
L_{\rm X} \simeq \frac{L_{\rm E}}{b}\left[1+\ln \left(
\frac{\dot{M}}{\dot{M}_{\rm E}} \right) \right],
\end{equation}
where $ b $ is the beaming factor.
\citet{k09} proposed an approximate formula 
\begin{equation}
b \simeq \frac{73}{\dot{m}^2}, 
\end{equation}
where $ \dot{m} = \dot{M} / \dot{M}_{\rm E}$. This formula is valid for $  \dot{m} \gtrsim 8.5$, otherwise  
the beaming effect is not operated (i.e., $ b = 1$). Accordingly, we assume that the probability of 
detecting a source along the beam is reduced by a factor of $ b $.

\section{Results and discussions}

In Figure~3 we plot the number distributions of the NS XRBs with a helium star (left panel) 
and a normal star donor (right panel) in the orbital period$-$mass transfer rate ($ P_{\rm orb}-\dot{M}_{\rm tr} $) plane. 
Only the systems with mass transfer rates larger than $10^{-8}\, M_{\odot}\rm\,yr^{-1}$ are presented. 
According to the mass transfer rates labelled on the left side axis of each panel, we can calculate the 
corresponding apparent X-ray luminosities by considering possible beaming effect, which are labelled 
on the right side axis for comparison. Note that the calculated X-ray luminosities between the left and right 
panels have a slight difference, as the Eddington limits for helium and hydrogen accretion are differently adopted.
The five black circles show the positions of the observed NS ULXs with known orbital periods
(see Table 1).
Each panel contains a $ 50\times 50 $ image matrix, the colors are scaled according to the number of the XRBs.
After recording the evolutionary tracks of  all NS binaries, we can calculate the number of binary systems 
passing through a specific matrix element by accumulating the product of their birthrates and the durations. 
We obtain that the NS XRBs with a helium star companion have the number of  $ \sim23 $ in a
Milky Way-like galaxy, the ages of such systems are typically 
 $\sim100 \rm Myr $. It can be seen that only a small group ($ \sim 9 $) of them  
can appear as ULXs because a high mass transfer rate of $\gtrsim 10^{-7}\, M_{\odot}\rm\,yr^{-1}$ is required
\citep{klk17}. When comparing these two diagrams, the calculated number distribution of 
the NS$-$normal star binaries seems to match the observations much better than the one of the NS$-$helium star binaries.
It should be noted that, however, the observed sample of the NS ULXs is still too small and subject to the observational 
bias that favoring luminous massive stars. 
In addition, the NSs in many ULX systems are likely to be unpulsed unless having high spin-up rates 
\citep{klk17}. It is unclear that whether the NS$-$helium star ULXs can be easily identified due to the 
emission of X-ray pulsations.

\begin{figure*}[hbtp]
\centering
\includegraphics[width=0.7\textwidth]{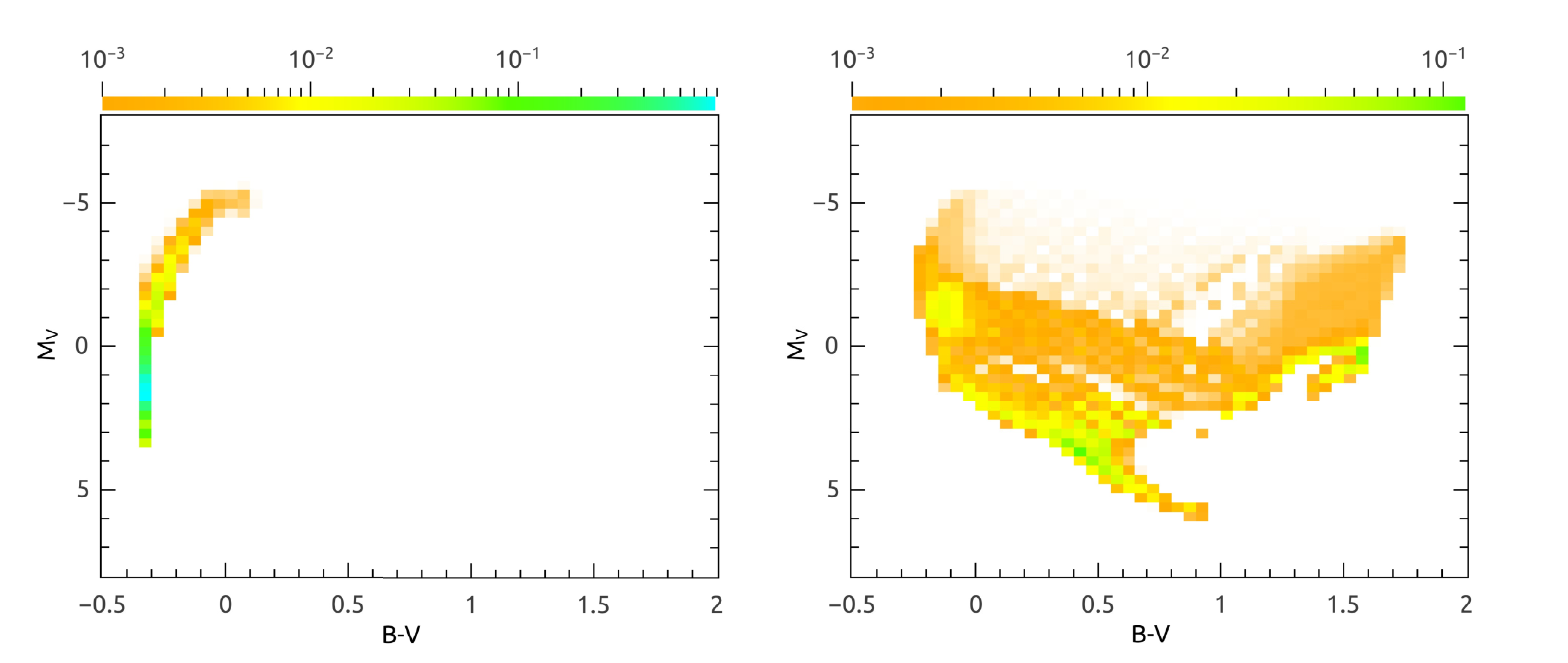}
\caption{Color-magnitude diagrams for the donors in the NS ULXs.
The left and right panels correspond to the distributions of the helium star and the normal star donors, respectively.
The colors are scaled according to the number of the ULX systems. 
   \label{figure1}}

\end{figure*}

\begin{figure*}[hbtp]
\centering
\includegraphics[width=0.7\textwidth]{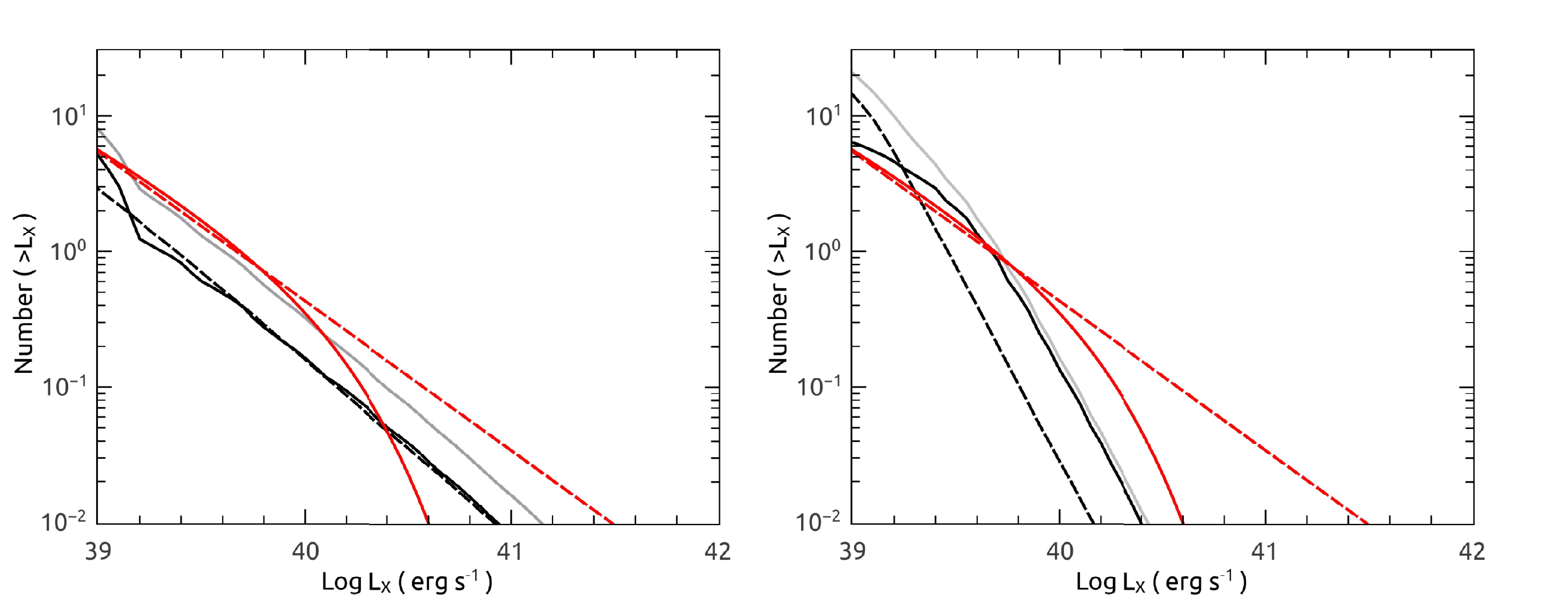}
\caption{Expected numbers of NS ULXs at the present epoch as a function of X-ray luminosity in 
Milky Way-like galaxies with a constant star formation rate of $ 3 M_{\odot}\,\rm yr^{-1} $. The
left and right panels correspond to the cases when adopting Equations (3) and (1) to calculate the
X-ray luminosity. In each panel, the black solid and black dashed curves respectively correspond to the ULX systems with 
a helium star and a normal star companion, and the grey solid curve corresponds to the whole population 
of the NS ULXs. Based on the observed ULX sample in nearby galaxies,
the luminosity functions are fitted when applying two models 
of a power-law with an exponential cut-off (red solid curve) and a pure power-law (red dashed curve). 
Here the fit parameters are taken from \citet{s11}.
   \label{figure1}}

\end{figure*}

Figure~4 depicts the number distributions of the NS ULXs with X-ray luminosities greater than 
$ 10^{39} \rm\, erg\,s^{-1} $ in the $M_{\rm d} - P_{\rm orb}$ plane. The effect of geometric beaming
on detection probability is taken into account. The left and right panels correspond 
to the NS ULXs with a helium star and a normal star \citep[see also][]{sl15} companion, respectively. 
Also plotted in Figure~5 is the histogram distributions of the ULX numbers 
as a function of the donor mass and the orbital period. 
The solid and dashed curves respectively correspond to the NS ULXs with a helium star and a normal star donor. 
We find that the majority of the ULX systems containing a helium star donor are close systems with orbital periods
distributing at a peak of $ \sim 0.1 $  day (with a tail up to $ \sim100 $ days), and the mass distribution of  the helium stars
has a maximum probability at $ \sim1M_{\odot} $ within the whole range of $\sim 0.6-2M_{\odot} $.  
This is apparent because the helium stars in short-period systems are less evolved and possess longer durations 
of mass transfer phases. The systems with a massive ($ \gtrsim2M_{\circ} $) helium star are hardly produced 
because of a combination of relatively low birthrates, short mass transfer 
durations and low detection possibilities (due to the beaming effect).

Figure 6 presents the color-magnitude diagrams for the donors of the NS ULXs.
The left and right panels correspond to the distributions of the helium star and the normal star donors, respectively.
In both cases, the majority of the ULX systems tend to have donors with absolute magnitude 
$ M_{\rm V} $ larger than $ -1 $ magnitude,
since the donors are predominantly low-mass ($ \lesssim 2M_{\odot} $) stars as mentioned above. 
If located in external galaxies, the donor counterparts are probably too dim to be detected. 
Furthermore the optical emission from the donor star in a ULX system
may be confused with that from the accretion disk around the compact star \citep{tf11}. 
Our results suggest that the characteristic of very short orbital 
periods can be used to distinguish the NS ULXs with a helium star donor.

In Figure~7 we show the X-ray luminosity function of the NS ULX population in 
Milky Way-like galaxies. The black solid and black dashed curves respectively correspond to 
the ULXs with a helium star and a normal star companion, and the grey solid curve corresponds 
to the whole population of the NS ULXs. The red curves present the fitted luminosity functions
of the observed ULX sample in nearby galaxies \citep[for details see][]{s11}. Note that these
fitted luminosity functions have been normalized to a star formation rate of $ 3 M_{\odot}\,\rm yr^{-1} $. 
In the left panel, the luminosity function for the ULX systems with a helium star donor has an 
obvious break at the position of $L_{\rm X} \sim 2\times 10^{39} \rm\, erg\,s^{-1} $, 
which corresponds to a point that the beaming effect starts operating. For the NS ULXs with a normal
star donor, the corresponding luminosity function also has a similar break but at $L_{\rm X} \sim 6\times 10^{38} 
\rm \,erg\,s^{-1} $, which is not covered in this diagram.
In the right panel, we show the X-ray luminosity function
without considering the beaming effect for comparison, by use of Equation~(1) to calculate the 
X-ray luminosity. We expect that the NS$-$helium star systems can contribute 
several ULXs in a Milky Way-like galaxy, which is comparable with that from the NS$-$normal star binaries. 
The number of the extremely luminous sources with $L_{\rm X} \geq 10^{40} \rm\, erg\,s^{-1} $
is only about $ \sim 0.3 $.  Although the ULX systems with a BH accretor are not included, 
our obtained number of the NS ULXs seems to match the observations.

Recently \citet{ws17} performed a BPS study on the origin of the ULX systems, the  
NS$-$helium star binaries were also included in their calculations. The NS$-$helium star binaries could 
only appear as the extremely luminous sources, and they
were expected to be very rare. At solar metallicity, the corresponding masses of the helium stars were 
in the range of $ 1.7 -  2.6M_{\odot}$ \citep{ws17}. 
This mass range is significantly larger than the one ($\sim 0.6-2.0M_{\odot}$) obtained by us, 
since we adopt the rotation-dependent (highly 
non-conservative) mass transfer model during the primordial binary evolution.  
This model allows a large amount of the primordial binaries to experience stable mass transfer,  
and then evolve to be relatively wide systems containing an NS and  an intermediate-mass ($ \sim3-8M_{\odot} $) 
normal star \citep{sl14}. 
The subsequent evolution of these wide binaries is expected to go into a CE phase when 
the intermediate-mass donor starts transferring mass to the NS.
After the CE evolution, the NS's companion may be a relatively low mass helium star. 
It is pointed out that a fraction of close binaries containing a (partially) recycled pulsar and a CO WD
(with mass of $ \sim0.6-1.3 M_{\odot} $) in the Milky Way are likely formed through this channel 
involving a CE phase \citep{t11}. If so, our results that involve relatively low mass helium stars 
can better reproduce the observed binaries containing a relatively light ($ \lesssim 1 M_\odot $) CO WD.

\section{Summary}

With a population synthesis study, we have shown that NS XRBs containing a helium star companion have a significant contribution to the ULX population in Milky Way-like galaxies. 
Assuming a constant star formation rate of $ 3 M_{\odot}\,\rm yr^{-1} $, we predict that there are several
NS$-$helium star ULX systems in a Milly Way-like galaxy, whose ages are typically $\sim 100\,\rm Myr $. 
These ULX systems favor short orbital periods, so their subsequent evolution will lead to the formation of 
close NS$-$WD binaries which are important gravitational wave sources.

\acknowledgements
We thank the referee for useful suggestions to improve this paper.
This work was supported by the Natural Science Foundation 
of China (Nos. 11973026, 11603010, 11773015, and 11563003)
and Project U1838201 supported by NSFC and CAS, and
the National Program on Key Research and 
Development Project (Grant No. 2016YFA0400803).


\clearpage

\end{document}